\documentclass[aps,prb,twocolumn,superscriptaddress,
preprintnumbers,amsmath,amssymb
,floatfix
,showpacs
]{revtex4}
\usepackage{graphicx}
\usepackage{dcolumn}
\usepackage{bm}

\begin{document}

\title{Metallization of Nanofilms in Strong
Adiabatic Electric Fields}

%% Notice placement of commas and superscripts and use of &
%% in the author list

\author{Maxim Durach}
\affiliation{
Department of Physics and Astronomy, Georgia State
University, Atlanta, Georgia 30303, USA}
\author{Anastasia Rusina}
\affiliation{
Department of Physics and Astronomy, Georgia State
University, Atlanta, Georgia 30303, USA}
\author{Matthias F. Kling}
\affiliation{Max-Planck-Institut f\"ur Quantenoptik,
Hans-Kopfermann-Stra{\ss}e 1, D-85748 Garching,
Germany}
\author{Mark I. Stockman}
\affiliation{
Department of Physics and Astronomy, Georgia State
University, Atlanta, Georgia 30303, USA}
\affiliation{Max-Planck-Institut f\"ur Quantenoptik,
Hans-Kopfermann-Stra{\ss}e 1, D-85748 Garching,
Germany}
\email{mstockman@gsu.edu}
\homepage{http://www.phy-astr.gsu.edu/stockman}

%\author{Maxim Durach$^{1}$, Anastasia Rusina$^{1}$, Mark I. Stockman$^{2,1}$}
 
%\begin{document}

\date{\today}

\begin{abstract}

We introduce an effect of metallization of dielectric
nanofilms by strong, adiabatically varying electric fields. The
metallization causes optical properties of a dielectric film to
become similar to those of a plasmonic metal (strong absorption and
negative permittivity at low optical frequencies).
%This is a quantum effect caused by
%crossing of the Fermi surface by localized electron states with an
%adiabatic transfer of the electron population across the nanofilm in
%real space.
The is a quantum effect,  which is  exponentially size-dependent,
occurring at fields on the order of $0.1 \mathrm{~V/\AA}$ and pulse
durations ranging from $\sim 1$ fs to $\sim 10$ ns for film
thickness $3-10$ nm.

\end{abstract}


\maketitle

%--------------------------------------------------------------------
\begin{figure}
\includegraphics[width=.42\textwidth]{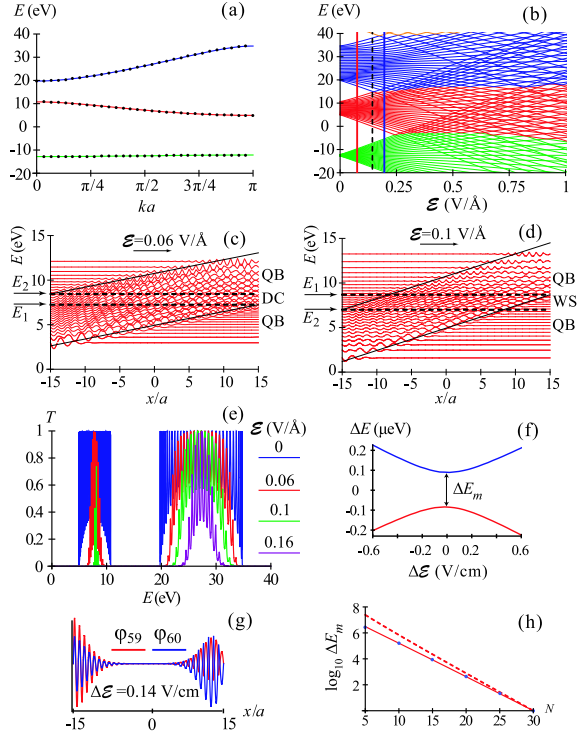}
\caption{
Field effect on electron states: spectrum, localization, transmission,
and mixing. The data are for an $N=30$ (9 nm) nanofilm thickness.
(a) Energy bands of infinite crystal (lines) and nanofilm (dots)
in zero electric field. Bands are color coded by the green (localized
band), red (valence band), and blue (conduction band). (b) Energy bands
of nanofilm as a function of the applied electric field (color coded as
above). (c) Valence band
levels in electric field $\mathcal E=0.06 \mathrm{~V/\AA}$.
The black lines and state notations are discussed in the text.
%The stopping
%points for carriers are shown by two slanted black solid lines as
%discussed in the text. The black dashed horizontal lines at energies
%$E_1$ and $E_2$ denote mobility edges. Quantum bouncers are denoted as
%QB, and delocalized conducting states are marked by DC.
(d) The same as
in panel (c) but for $\mathcal E=0.1 \mathrm{~V/\AA}$.
%The Wannier-Stark
%states are denoted by WS.
(e) Transmission coefficient $T$ as a
function of electron energy $E$ for different applied fields (color
coded as shown). (f) Anticrossing between the lowest-energy level in
the conduction band (blue) and the highest-energy level of the valence
band (red). Energy $\Delta E$ and field $\Delta \mathcal E$ are given
with respect to the crossing point. (g) The wave functions of the 
anticrossing states of panel (f) with the corresponding color coding for
$\Delta \mathcal E=0.14 \mathrm{~V/cm}$.
(h) The minimum splitting of the anticrossing levels as a function
of the film thickness expressed as $N$.
The black dots are obtained
from numerical computation. The red line is calculated
using Eq.\ (\ref{Delta_E_m}) and the dashed blue line from
Eq.\ (\ref{Delta_E_m_1}).}
\label{moderate_fields}
\end{figure}
%--------------------------------------------------------------------

Effects of strong electric fields on electron states in crystals have
attracted a great deal of attention over many decades going back to
Zener who predicted breakdown due to interband tunneling 
\cite{Zener_Proc_Royal_Soc_1934_Breakdown}. In insulators this requires
electric fields on the order of atomic fields $\mathcal E\sim 1-10$
V/\protect{\AA}. Interest to strong-field condensed matter physics has
recently greatly increased due to the availability of such strong
electric fields in laser pulses of intensities $I\sim 10^{13}-10^{15}
\mathrm{~ W/cm^2}$. Ultrashort laser pulses
% of duration on the order of 
with a few optical oscillations
\cite{Corkum_Krausz_Nature_Physics_2007_Attosecond_Science,%
Krausz_Ivanov_RevModPhys.81.163_2009_Attosecond_Review} open up a
possibility to study ultrastrong field phenomena in solids during periods
of time too short for the lattice ions to move
significantly.
%from their positions, thus preserving the lattice integrity. 
Recent \emph{ab initio} calculations
\cite{Otobe_et_al_PRB_2008_Metallization_in_Diamond}
%Otobe_Yabana_Iwata_J_Phys_Cond_Mat_2009_Si_Metallisation} 
% using local density approximation 
have reproduced
% the electron transfer to the conduction band 
the Zener breakdown in insulators induced by a laser pulse of intensity
$\sim 10^{15} \mathrm{~W/cm^2}$.
%, which is consistent with the Zener breakdown. 
Other strong-field phenomena that can be observable in
crystals at a comparable field strength are the appearance of localized
electron states, Wannier-Stark ladder in the energy spectrum
\cite{Wannier_PR_1960_Wannier_States_in_Strong_Fields,%
Shockley_PRL_1972_Wannier_Stark_Ladder}, and 
Bloch oscillations.
\cite{Bloch_Z_Phys_1929_Functions_Oscillations_in_Crystals} 
At orders of magnitude lower intensities, low-frequency
%(quasistationary) 
optical fields cause a reduction of the band gap in
semiconductors and insulators (Franz-Keldysh effect, FKE)
\cite{Franz_Nat_Forsch_1934_Absorptionkante,%
Keldysh_JETP_1958_Franz-Keldysh_Effect}. The quantum confined
FKE  takes place in semiconductor quantum wells
and is determined not by the field but by the total potential drop.
\cite{Miller_Chemla_Schmitt-Rink_PRB_1986_Quantum-Confined_Franz-Keldysh_Effect}
It requires typical fields $\mathcal E\sim 10^{-3} \mathrm{~V/\AA}$.
% i.e., they are orders of magnitude less the fields needed for efficient Zener breakdown.

In this Letter we introduce an effect of metallization in insulator
nanofilms, which is predicted to occur in applied electric fields
$\mathcal E\sim 0.1 \mathrm{~V/\AA}$. It is based on adiabatic electron
transfer in space across the nanofilm. The minimum duration of the field
pulse required for the adiabaticity exponentially depends on
the crystal thickness varying from $\sim 1$ fs for a 3 nm film to
$\sim 10$ ns for a 10 nm film thickness. This metallization effect
manifests itself by a dramatic change in the optical properties of the
system, which start to remind those of metals. In particular, 
%low-frequency absorption appears along with 
plasmonic phenomena emerge. 

To demonstrate the metallization effect, we need to solve the
one-electron Schr\"odinger equation for a periodic potential plus a
uniform electric field very accurately. We will employ the widely used
Kronig-Penney model for electrons in a film confined in the $x$
direction by an infinite potential well. The corresponding potential
energy (neglecting the electron-electron interaction) is
\begin{equation}
V(\mathbf{r})=\begin{cases}
U(x)+U(y)+U(z)+ e \mathcal{E}x &
|x|<L/2\\
\infty & |x|\ge L/2
\end{cases}
\label{potential}
\end{equation}
where $U(x)=-\alpha \sum_{n=-\infty}^\infty \delta(x-na)$, and $a$ is
the lattice constant. Crystal thickness $L$ is determined by the number
of the lattice periods $N$ in the $x$ direction, $L=Na$.
Though this model does not precisely correspond to
any real system, it is exactly solvable and catches the qualitative
features of the strong-field phenomena from formation of the quantum
bouncer (QB) states  to band gap collapse  and metallization transition.
% Wannier-Stark localization and ladder formation.
%
We use the transfer matrix to find an exact
solution of the Schr\"odinger equation with potential (\ref{potential})
-- see Sec. II  of Supplemental Material\cite{EPAPS}.
% where specific
%expressions for the wave functions and the transfer matrix are given in
The zero boundary conditions for the wave function at
$z=\pm L/2$ have been imposed and energies found 
semi-analytically using the shooting method.

The energy bands of the infinite crystal in the zero field are shown in
Fig.\ \ref{moderate_fields} (a) by  color coded lines for
the three lowest bands:  localized (green), valence
(red), and conduction (blue). The discrete electronic levels
for the $N=30$ film in zero field are represented by the black dots
superimposed on these infinite-crystal dispersion curves.
Our computations are
made for $\alpha=13.5~\mathrm{eV}~\mathrm{\AA}$ and $a=2.7~\mathrm{\AA}$
to result in the gap between the conduction and
valence bands to be $E_g=9~\mathrm{eV}$, which is the same as for
$\mathrm{SiO}_2$. We consider an insulator (semiconductor) film where
the localized and valence bands are completely filled, and the
conduction band is empty. This corresponds to a population of 16
electrons per unit cell (taking spin into account)
and a reasonable electron density of $n = 8 \cdot
10^{23}~\mathrm{cm}^{-3}$.

The energy spectrum of the nanofilm as a function of the applied field
$\mathcal E$ normal to the film is
shown in Fig.\ \ref{moderate_fields} (b) where the filled bands are
coded by the green (localized band) and red (valence band), and the
empty conduction band is indicated by blue. With an increase of
$\mathcal E$, the linear Stark effect takes place,
and the conduction-valence band gap decreases,
completely closing at a metallization field $\mathcal E_m= 0.148
\mathrm{~V/\protect{\AA}}$ indicated by the vertical dashed line. 
This behavior can be understood from analytical theory presented in
Sec.\ V of the Supplemental Material\cite{EPAPS}, where Eq.\ (38) expresses
$\mathcal{E}_m$ as
\begin{equation}
\mathcal E_m=\frac{E_g}{e\left[L-\zeta_1\Lambda_e(\mathcal E_m)-%
\xi_1\Lambda_h(\mathcal E_m)\right]}~.
\label{E_m}
\end{equation}
Here an electric-field quantum confinement length is
$\Lambda_{e,h}(\mathcal E)=
[\hbar^2/(2 m^\ast_{e,h}e\mathcal E)]^{1/3}$, where $m^\ast_{e,h}$ are
the effective masses for electrons and holes, and $\xi_1$ and $\zeta_1$ are
the first roots of $\mathrm{Ai}(-x)$ and $\mathrm{Ai}^\prime(-x)$,
respectively.

The band gap can be estimated
as $E_g\sim \pi^2\hbar^2/(2ma^2)$. From this, we can
estimate 
$\Lambda_{e,h}\sim \left[m/(\pi^2 m^\ast_{e,h})a^2 L\right]^{1/3}\ll L$.
Neglecting $\Lambda_{e,h}$ in comparison with $L$, one obtains a
very good approximation for $\mathcal E_m$, band edges $E_{b,t}$ (where
$b$ and $t$ stand for the top and bottom),
and band gap $E_g$ as 
\begin{equation}
\mathcal E_m=\frac{E_g}{eL}~,~
E_{b,t}(\mathcal E)= E_{b,t}\mp e \mathcal{E} \frac{L}{2}~,~
E_g(\mathcal E)= E_g-e \mathcal{E} L~.
\label{egdes}
\end{equation}
This implies a linear Stark effect near the metallization point,
in an excellent agreement with Fig.\
\ref{moderate_fields} (b).

%An electric field applied normally to the nanofilm has a pronounced
%effect on the carriers as illustrated in Figs.\
%\ref{moderate_fields} (c) and (d) where the wave functions for the
%valence band are displayed. The states at the bottom of the valence band
%have a positive effective mass $m_e^\ast>0$ [cf.\ panel (a)] and behave
%as electrons in the applied electric field, while the states at the top
%of this band possess a negative effective mass $-m_h^\ast$ and,
%consequently, behave as holes. The electric field tend to localize these
%electron and hole states at the opposite boundaries of the nanofilm.

An applied normal electric field causes the appearance of states
localized between the corresponding boundary at $x=\pm L/2$ and the
stopping points whose coordinates $x_s$ and energy $E$ are related as
$E=E_{b,t}+e \mathcal E x_s$ [cf.\ Eqs.\ (29) and (33) of the
Supplemental Material\cite{EPAPS}]. These relations are represented by the slanted
black lines in Figs.\ \ref{moderate_fields} (c) and (d). 
Carriers at the conduction band bottom behave as electrons, and
those at the top of the valence band behave as holes. The corresponding
mobility edges are denoted as $E_1$ and $E_2$. For moderate fields
[panel (c)], the states with energies $E>E_2$ or $E<E_1$ are
Bloch-electron quantum bouncers (QBs)
\cite{%Gea-Banacloche_AJP_199_Quantum_Bouncer,
Goodmanson_AJP_2000_Quantum_Bouncer}. The states with intermediate
energies $E_2>E>E_1$ do not have stopping points and are delocalized,
conducting (DC).

For a stronger electric field, as shown in Fig.\
\ref{moderate_fields} (d), the mobility edges overlap, $E_1>E_2$,
and the DC states disappear. Instead, localized Wannier-Stark (WS)
states appear with energies $E_1>E>E_2$. These states are very close to
those in infinite lattices.
\cite{Shockley_PRL_1972_Wannier_Stark_Ladder}
Their contribution to
the {\it static} conductivity vanishes -- see the next paragraph.

Assume that the barriers at the boundaries of the nanofilm are 
transparent enough to allow for tunneling through. Then it is physically
meaningful to find the transmission coefficient  $T$ for carriers injected at
a certain energy, which is plotted in
Fig.\ \ref{moderate_fields} (e) for the valence and conduction bands
for different values of $\mathcal{E}$. As $\mathcal{E}$
increases, the transmission band collapses along with the disappearance
of the delocalized conducting (DC) states.

Metallic behavior is actually characterized by two different phenomena:
dc conductivity and negative $\mathrm{Re}\,\varepsilon$ (where
$\varepsilon$ is permittivity) that contributes to plasmonic phenomena.
They do not necessarily both take place simultaneously. We have shown
above that the field-induced localization eliminates the dc
conductivity.

Now we will show that, to the opposite, the applied field that increases
slowly (adiabatically) turns an insulator crystal into a metal
optically, which we call the metallization effect. A distinct property
of metals is the absence of a band gap around the Fermi energy. The two
levels at the edges of the valence and conduction band at field
$\mathcal E_m$ experience an anticrossing as Fig.\
\ref{moderate_fields} (f) shows. The band gap 
%(splitting between these two levels) 
is very small, $\Delta E_m\sim 0.1 \mathrm{~\mu eV}$, which
%This smallness 
stems from a very little overlap between the wave
functions of the two edge state QBs 
%, which are displayed in 
[Fig.\ \ref{moderate_fields} (g)].
% at the anticrossing point.

The minimum splitting of these anticrossing levels is related to  the
matrix element of the Zener-type tunneling between the valence and
conduction bands. This splitting can be calculated analytically using
the quasiclassical approximation of Refs.\
\onlinecite{Kane_J_Phys_Chem_Solids_1959_Zener_Tunneling,%
McAfee_Ryder_Shockley_Sparks_PR_1951_Zener_Current_in_Ge_pn_Junctions,%
Glutsch_PhysRevB.69.235317_2004_Zener_Tunneling} and   Eq.\ (\ref{egdes}).
We obtain
\begin{eqnarray}
\Delta E_m&\propto& \sqrt{\mathcal E_m} \exp\left[-\pi %
\sqrt{\mu E_g^3}/(4\hbar e\mathcal E_m)\right]~
\label{Delta_E_m}\\
&\propto& \exp\left[-\pi %
\sqrt{\mu E_g}L/(4\hbar)\right]/\sqrt{L}~,
\label{Delta_E_m_1}
\end{eqnarray}
where the reduced mass is $\mu=m^\ast_e m^\ast_h/(m^\ast_e + m^\ast_h)$.
% Eq.\ (\ref{Delta_E_m_1}) is obtained using an approximate value
% $\mathcal E_m$, Eq.\ (\ref{egdes}). 

The dependence of the minimum band splitting $\Delta E_m$ 
on $N=L/a$ from Eq.\ (\ref{Delta_E_m}), where the critical
field is given by Eq.\ (\ref{E_m}), is shown in Fig.\
\ref{moderate_fields} (h) with a solid red line. It is in an
excellent agreement with numerically computed points obtained from our
quantum-mechanical solution, which are displayed as bold dots. Dependence
given by a simplified expression (\ref{Delta_E_m_1}) is shown by the
dash blue line in Fig.\ \ref{moderate_fields} (h) and is
a good approximation. 

\begin{figure}
\includegraphics[width=.42\textwidth]{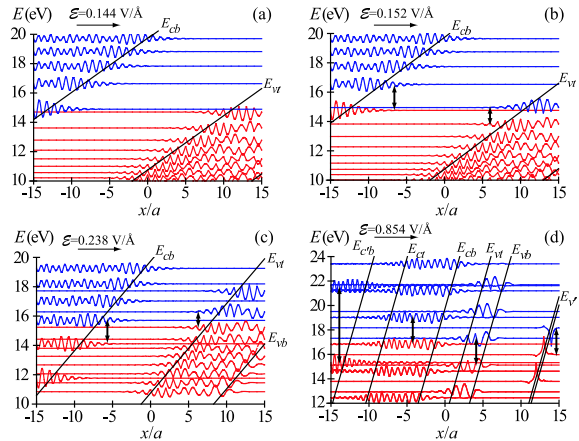}
\caption{\label{multizone_levels_30}
Electron states in metallizing fields for $N=30$. The occupied states
(below Fermi energy $E_F$) are indicated by red and the vacant ones by blue.
(a) Electrons in electric field
$\mathcal E=0.144 \mathrm{~V/\AA}$ slightly less than the metallization
threshold $\mathcal E_m=0.148  \mathrm{~V/\AA}$. (b) Electrons in electric field
$\mathcal E=0.152 \mathrm{~V/\AA}$ slightly exceeding
$\mathcal E_m$. The black double arrows denote the dominant low-frequency
optical transitions. (c) and (d) The same as for previous panels but for fields
$\mathcal E=0.238 \mathrm{~V/\AA}$ and $\mathcal E=0.854
\mathrm{~V/\AA}$, correspondingly.
}
\end{figure}

\begin{figure}
\includegraphics[width=.42\textwidth]{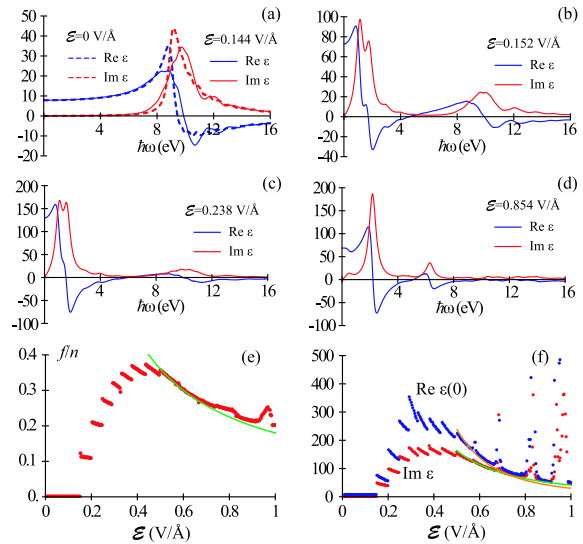}
\caption{\label{spectra_permittivity}
Optical properties of nanofilms with $N=30$ in electric fields.
(a) Below metallization threshold:
real (blue) and imaginary (red) parts of permittivity
$\varepsilon$ for fields $\mathcal E<\mathcal E_m$.
The dashed curves correspond to the zero field, while the solid curves are for
$\mathcal E= 0.144 \mathrm{~V/\AA}$. (b) Above metallization
threshold: $\mathrm{Re}\,\varepsilon$ and $\mathrm{Im}\,\varepsilon$ 
for $\mathcal E= 0.152 \mathrm{~V/\AA}$. Two low-frequency spectral 
peaks correspond to the intraband transitions between
the QBs. (c) The same as in the previous panel but in a stronger field
$\mathcal E=0.238 \mathrm{~V/\AA}$. (d) The same as (c) but for
$\mathcal E=0.854 \mathrm{~V/\AA}$. The three low-frequency absorption peaks
correspond to FKE, intraband transitions between WS states, and
intraband transition between QBs, respectively, listed in the
order of increasing transition frequency.
(e) Combined oscillator strength of low-frequency transitions
as a function of applied electric field $\mathcal E$. The green line is
an analytical approximation -- see text. 
(f) Real part of dielectric permittivity at zero frequency (blue)
and imaginary part (red) of dielectric function averaged over
low-frequency spectrum ($E<3$ eV) as functions of the applied field.
Analytical approximations (green and orange lines) are
described in text.
} 
\end{figure}

Now we turn to the metallization. Consider first field $\mathcal
E=0.144 \mathrm{~V/\AA}$, which is slightly less than $\mathcal E_m=0.148
\mathrm{~V/\AA}$. The corresponding energy levels near the Fermi energy
$E_F\approx 14.8$ eV are shown in Fig.\ \ref{multizone_levels_30}
(a). We assume temperature to be sufficiently low so only the states
below $E_F$ are occupied (shown by red) and those with $E>E_F$ are
vacant (blue color). The valence band states are QBs at the right
boundary, and those of the conduction band are QBs at the left boundary.
Though the band gap is very small, the low-frequency transitions between
these two bands are drastically suppressed due to the very weak spatial
overlap. This agrees with the corresponding optical spectra for $\mathcal
E=0.144 \mathrm{~V/\AA}$ shown in
Fig.\ \ref{spectra_permittivity} (a), which are almost the same as
for $\mathcal E=0$.

%--------------------------------------------------------------------
%--------------------------------------------------------------------

%--------------------------------------------------------------------
%--------------------------------------------------------------------

Assume that the field is {\it slowly} increased to $\mathcal E>\mathcal
E_m$, so that the level anticrossing shown in Fig.\
\ref{moderate_fields} (f) is {\it adiabatically} passed, which
requires that the passage time $t_p\gtrsim \hbar/\Delta E_m$
(cf.\ Landau-Zener theory \cite{Zener_Proc_Royal_Soc_1932_Diabatic_Crossing}).
This 
necessitates that the nanofilm is thin enough -- cf.\ Eq.\
(\ref{Delta_E_m_1}). Then the passage is adiabatic, the system persists in
the ground state, and the electron population remains below the Fermi surface
-- see Fig.\ \ref{multizone_levels_30} (b).
%
%The upper
%QB level originating
%from the valence band has crossed the Fermi surface and is vacant, and
%the lower  QB from the conduction band is now below the
%Fermi energy and is filled.
%Thus a
%
At each boundary of the film, there are
QBs on the opposite sides of the Fermi surface, which
significantly overlap in space. This allows for strong electron transitions
at low frequencies as shown by the vertical double arrows.
Note that the transition probability between two QBs rapidly decreases
with the transition frequency $\propto \omega^{-4}$ [see Eqs.\ (30) and
(34) of Supplementary Material\cite{EPAPS}]. Correspondingly, the optical
absorption is dramatically shifted to the red and infrared (ir) parts of the
spectrum as displayed in Fig.\ \ref{spectra_permittivity} (b).
In the red spectral region,
$\mathrm{Re}\,\varepsilon<0$,  which
is characteristic of metals. This is the metallization effect.

With a further adiabatic increase of the field to $\mathcal E=0.238
\mathrm{~V/\AA}$, more levels cross the Fermi surface -- see 
Fig.\ \ref{multizone_levels_30} (c). A stronger low-frequency
absorption takes place [Fig.\
\ref{spectra_permittivity} (c)].  This signifies a more developed
metallization.

%Note that at this field there are WS states formed in the
%valence band with $E<E_{vb}$, but these states do not yet contribute to
%the optical absorption.

For a very strong field case illustrated in Fig.\
\ref{multizone_levels_30} (d), the Fermi surface separates levels
originating from four initial energy bands. The dominating transitions
are those between WS states, and their frequency is $\approx e\mathcal E
a$ increasing linearly with $\mathcal E$. This leads to a general shift
of the optical spectra to the blue -- see Fig.\
\ref{spectra_permittivity} (d) where the absorption maximum is now
at approximately 2 eV.

%Note that the small
%peak at the low frequencies ($<1$ eV) is due to the transition between
%the nearest levels separated by the Fermi surface and originating from
%two different bands (valence and conduction). It is quite analogous
%to the FKE \cite{Franz_Nat_Forsch_1934_Absorptionkante,%
%Keldysh_JETP_1958_Franz-Keldysh_Effect}.

To emphasize the metallization transition, we show in Fig.\
\ref{spectra_permittivity} (e) the oscillator strength $f$ of the
low frequency (below 3 eV) transitions as a function of $\mathcal E$.
Note that $f/n\approx 1$ is characteristic of metals. This figure demonstrates
a resemblance between the metallization and a quantum phase transition.
Up to the metallization critical field $\mathcal E_m=0.148
\mathrm{~V/\AA}$, the oscillator strength $f$ is practically zero. After
that, $f$ increases in steps, each corresponding to a pair of QBs
adiabatically crossing the Fermi surface. At the maximum, $f/n\approx
0.4$ implying that $\approx 40\%$ of the total number of electrons
contributes to this metallic behavior. A further increase of $\mathcal
E$ leads to the WS states crossing the Fermi surface and the
oscillator strength decreasing [cf.\ Eq.\ (46) of the
Supplemental Material\cite{EPAPS}] $f\propto \mathcal E^{-1}$ as shown by the
solid green line.

Behavior of the permittivity across the metallization transition is
displayed in Fig.\ \ref{spectra_permittivity} (f). Below $\mathcal
E_m$, $\varepsilon$ is relatively low. For $\mathcal E>\mathcal E_m$,
both $\mathrm{Re}\,\varepsilon(0)$ and $\mathrm{Im}\,\varepsilon$ increase
in steps due to the QB states crossing the Fermi surface. After reaching
the maximum, the main contribution to $\varepsilon$ shifts to the
transitions between WS states originating from the same band (either
valence or conduction). This causes decrease in $\varepsilon$ as
shown by the green and orange lines computed using Eqs.\ (47)-(48) of
the Supplemental Material\cite{EPAPS}. As field increases, there are also
transitions between the WS states originating from the different bands,
% These transitions are due to photon-assisted tunneling and
leading to the sharp peaks in Fig.\ \ref{spectra_permittivity} (f). 

Concluding (see also Sec. VIII of Supplemental Material\cite{EPAPS}), we have predicted an effect of metallization in dielectric nanofilms induced by an adiabatically increasing applied field. The localized states crossing the Fermi surface cause optical absorption
extending from very low (THz) frequencies over all optical
region. In the near-ir and red spectral region, $\mathrm{Re}\,
\varepsilon<0$ is predicted. This property is characteristic of
metals and allow for a multitude of nanoplasmonic effects.
%
% Counterintuitevely, metallization
%occurs on the same range of intensities as the field-induced electron
%localization, which leads to the disappearance of the intraband
%transmission and vanishing dc conductivity.
%
The metallization cardinally differs from the Zener breakdown in bulk
crystals, which is clear from much lower fields required
($\mathcal E_m\sim 0.1 \mathrm{~V/\AA}$ for $L=10$ nm).
%in contrast to $\mathcal E\gtrsim 1-10 \mathrm{~V/\AA}$ for the Zener breakdown). 
In fact, the metallization is defined not by the field $\mathcal{E}$ per se but by the total potential difference $\Delta U=\mathcal{E}_m L=E_g/e$.
Due to the requirement of adiabatic passage to the metallized state, the
rise time $t_p$ of the applied electric field {\it exponentially}
increases with $L\sqrt{\mu E_g}/\hbar$. For instance, for the considered case ($L=10$ nm, $E_g=9$ eV),  $t_p \gtrsim 10$ ns, while for $L=3$ nm the passage is much faster:  $t_p\gtrsim 1$ fs.

The manifestations of the metallization depend on the way the field is induced in the nanostructure. For the excitation by an optical or THz wave electric field, the metallization will cause  high values of the permittivity and, consequently, bring about the plasmonic behavior of the system. This will lead, in particular, to screening of the external fields limiting the internal fields to $\mathcal{E}\sim \mathcal{E}_m$. Note that $\mathcal E_m\sim 0.1 \mathrm{~V/\AA}$ corresponds to the wave intensity $W\sim 10^{11} - 10^{12} \mathrm{~W/cm^2}$, which is well tolerated by nanostructured plasmonic metals -- cf.\ Ref.\ \onlinecite{Kim_et_al_Nature_2008_Nanoplasmonic_EUV}. In this case, 
the metallization effect is completely reversible. This will open up the field of nanoplasmonics to a variety of new dielectric and semiconductor nanosystems with a plethora of new phenomena possible. Among potential applications, is an ultrafast field-effect transistor where an ir or optical
fs pulse controls a dielectric gate.
%
%In particular, this will be highly sensitive to the carrier-envelope phase, which can find %applications in ultrafast science %\cite{Corkum_Krausz_Nature_Physics_2007_Attosecond_Science,%
%Krausz_Ivanov_RevModPhys.81.163_2009_Attosecond_Review}.

 In contrast, for dc- to microwave-frequency potential applied via electrodes,  the external potential difference is fixed.  Then the  metallization will lead to and is a new mechanism of the dielectric breakdown, which is fundamentally different from both the Zener and avalanche mechanisms. Such a situation is characteristic for the nanometric layers of the  insulator in field-effect transistors and super-capacitors, with far-ranging technological ramifications for microelectronics and energy storage.

%
%This
%is due to the fact that the metallization occurs when an electron
%acquires a band gap energy on the distance of the film thickness
%($e\mathcal E_m L \sim E_g$) while the Zener tunneling is strong when
%the same energy is acquired at the lattice period ($e\mathcal E_m a \sim
%E_g$). 
%
%This makes the metallization
%essentially a nanoscopic effect limited to a size $\propto (\mu
%E_g)^{-1/2}$.
%
%Thus the metallization is a very strongly size-dependent nanometric
%effect that becomes ultrafast for $L$ on the order of a few nanometers.
%As the dependence on the band gap $E_g$ suggests, the metallization in
%conventional semiconductors occurs for lower and faster fields than in
%wide-band semiconductors or dielectrics considered in this Letter.

We appreciate %useful 
discussions with F. Krausz and R. Ernstorfer. This
work was supported by grants from the Chemical Sciences, Biosciences and
Geosciences Division of the BES Office of the
% Basic Energy Sciences, 
% Office of Science, 
US Department of Energy, a grant CHE-0507147 from NSF, 
and
% a grant
the US-Israel BSF. 

%\bibliography{../../texbib/references}

\end{document}